\documentclass[pdflatex,sn-nature]{sn-jnl}

\usepackage{graphicx}%
\usepackage{multirow}%
\usepackage{amsmath,amssymb,amsfonts}%
\usepackage{amsthm}%
\usepackage{mathrsfs}%
\usepackage[title]{appendix}%
\usepackage{xcolor}%
\usepackage{textcomp}%
\usepackage{manyfoot}%
\usepackage{booktabs}%
\usepackage{algorithm}%
\usepackage{algorithmicx}%
\usepackage{algpseudocode}%
\usepackage{listings}%

\theoremstyle{thmstyleone}%
\theoremstyle{thmstyletwo}%

\theoremstyle{thmstylethree}%

\begin{document}

\title[Article Title]{Mask-free fast patterning of organic light-emitting diode pixels using laser-assisted close-space sublimation
}




\author*[1]{\fnm{Subhamoy} \sur{Sahoo}}
\equalcont{Presently at Lehrstuhl für Technische Physik, Julius-Maximilians-Universität Würzburg, Am Hubland, 97074, Würzburg, Germany}\email{subhamoysahoo@gmail.com}

\author[2]{\fnm{Jain} \sur{Jose}}

\author[2]{\fnm{Mani} \sur{R}}

\author[2]{\fnm{Arghya} \sur{Saha}}

\author[2]{\fnm{Kanimozhi} \sur{V}}

\author[2]{\fnm{Dhruvajyoti} \sur{Barah}}

\author[2]{\fnm{R. Bairava} \sur{Ganesh}}

\author[3]{\fnm{Amitava} \sur{Majumdar}}

\author[1]{\fnm{Jayeeta} \sur{Bhattacharyya}}

\author[2]{\fnm{G} \sur{Rajeswaran}}

\author*[2]{\fnm{Debdutta} \sur{Ray}} \email{dray@ee.iitm.ac.in}

\affil[1]{\orgdiv{Department of Physics}, \orgname{Indian Institute of Technology Madras}, \orgaddress{\city{Chennai}, \postcode{600036}, \country{India}}}

\affil[2]{\orgdiv{AMOLED Research Center, Department of Electrical Engineering}, \orgname{Indian Institute of Technology Madras}, \orgaddress{\city{Chennai}, \postcode{600036}, \country{India}}}

\affil[3]{ \orgname{Grantwood Technology Pvt. Ltd.}, \orgaddress{\city{New Delhi}, \postcode{110019}, \country{India}}}






\abstract{Existing patterning processes for organic light-emitting diode displays offer micrometer-scale precision but are constrained by long processing times for large-area substrates.
In this work, we study a fast growth method for patterned organic film deposition, aimed at applications including active-matrix organic light-emitting diode displays. The approach employs a specially engineered donor substrate in a close-space sublimation configuration combined with laser heating. The donor substrate incorporates spatially patterned absorber and reflector layers that enable selective, one-step or two-step transfer of organic material onto a receiver substrate. We analyze the optical response and heat-transfer dynamics that govern the selective transfer mechanism and demonstrate precise pixel patterning with micrometer-scale spatial fidelity. The reliability and practical applicability of the method are validated by fabricating light-emitting diode devices by using this rapid transfer process and benchmarking their performance against devices produced via conventional vacuum thermal evaporation. The resulting devices exhibit comparable optoelectronic performance, confirming the robustness and technological relevance of the proposed strategy for scalable fabrication of patterned organic light-emitting diodes for display application.
}

\keywords{Mask-free pixel patterning, Laser-assisted CSS, OLED, micro-pixels, display}



\maketitle

\section{Introduction}\label{sec1}
Organic Light Emitting Diode (OLED) based displays have found commercial success in displays due to their high contrast, excellent color rendering, wide view angle, and power efficiency \cite{tsujimura2017oled, geffroy2006organic, burrows1997achieving, forrest2004path, matsusue2005charge, king2009exploiting, yang2015recent, kalyani2017principles}. The current generation OLED pixels in mobile phone application is dominantly fabricated using linear evaporation sources for material transfer and fine metal mask technology for defining the pixel \cite{kim2020fine}. The linear evaporation method offers a major improvement over point sources due to homogenous layer deposition over a large area substrates \cite{van200227, hamer200569}. This decreases the \textcolor{black}{total accumulated cycle time (TACT)} for fabrication of OLED display units. The linear source has better material utilization efficiency compared to point sources. However, their operation is based on the classical vapor thermal evaporation (VTE) technique, and film growth rates are limited to sub-nm per second \cite{van200227, hamer200569}. While the material utilization efficiency for linear sources is better than a point source, there is still around 50\% material wastage incurred during the growth process.

There have been a number of processes developed to speed up the deposition process. One such process is \textcolor{black}{close-space sublimation} (CSS) \cite{tam2023low}. Unlike VTE, where a large source-substrate separation is required for uniform deposition, in CSS the source and substrate are placed close to each other with a separation of 1-5 mm. The deposition rates are considerably higher ($>$10 nm/s) for CSS which has been demonstrated for CdTe and perovskite-based  solar cells \cite{ferekides2000high, jiang2009growth, sharmin2023effect, rodkey2024close, diercks2026close}. The source used  for CSS is a pellet, which is prepared either by pressing powder under high pressure or using a powder bed. A modified version of CSS utilizes the transfer of pre-coated films from a planar donor substrate to a receiver substrate using Joule heating or energy transfer from a laser under vacuum. In CSS, single as well as mixed layers can be grown by transferring the film using Joule heating or flash heating at reduced
temperatures compared to that of VTE from powder sources \cite{tam2023low, tam201912}. However, the molecular weights of the constituent molecules should be similar to transfer the mixed films using Joule heating. In addition to fast transfer, CSS can be used to form patterned films. In Laser-Induced Forward Transfer (LIFT), a layer of the material which needs to be transferred is deposited on a transparent substrate, following which irradiation from laser pulses through the substrate are absorbed by the film leading to its heating and subsequent ablation to a receiver substrate \cite{fardel2007fabrication, shaw2013optimisation}. Materials may degrade under direct irradiation, and variations were proposed to reduce degradation.

In laser-induced thermal imaging (LITI), an integrated light-to-heat conversion (LTHC) layer absorbs laser irradiation, generating localized heat that facilitates the release and transfer of the overlying material to the receiver substrate \cite{cho2012enhanced, wolk200836, suh2003enhanced}. Both the processes can be used to create patterned deposition, which is required in AMOLED displays. Whereas the
state-of-the-art smartphone displays are fabricated using patterning based on Fine Metal Mask (FMM) technology \cite{kim2020fine}. The pattern accuracy in LITI can be a few micrometers and is an improvement over current generation FMM technology, where the pattern accuracy is in the order of 15-20 $\mu$m \cite{kim202061, kim201927}.  However, both LIFT and LITI rely on a single laser head to write multiple lines, limiting throughput for large-area substrates.

In this work, we study a fast growth method where patterned films can be grown for organic devices like AMOLED displays using a specially patterned donor substrate based on the CSS with laser heating \cite{Rajeswaran2015}. The donor substrate is composed of patterned light absorber and reflector layers, which are used to selectively transfer films to a receiver substrate in one step or two steps. We explore the optical properties and heat flow dynamics of the substrate, leading to the selective transfer process. The reliability of the proposed method was subsequently assessed by benchmarking the performance of LEDs fabricated using this fast transfer process against devices realized using conventional vacuum thermal evaporation (VTE). The resulting devices exhibited comparable optoelectronic characteristics, thereby substantiating the robustness and practical viability of the strategy. In addition, the accuracy of pixel patterning on the substrate was quantitatively evaluated, demonstrating high spatial fidelity enabled by this method.


\section{Results}\label{sec2}
\begin{figure}[ht!]
	\centering	
	\includegraphics[width=0.8\textwidth]{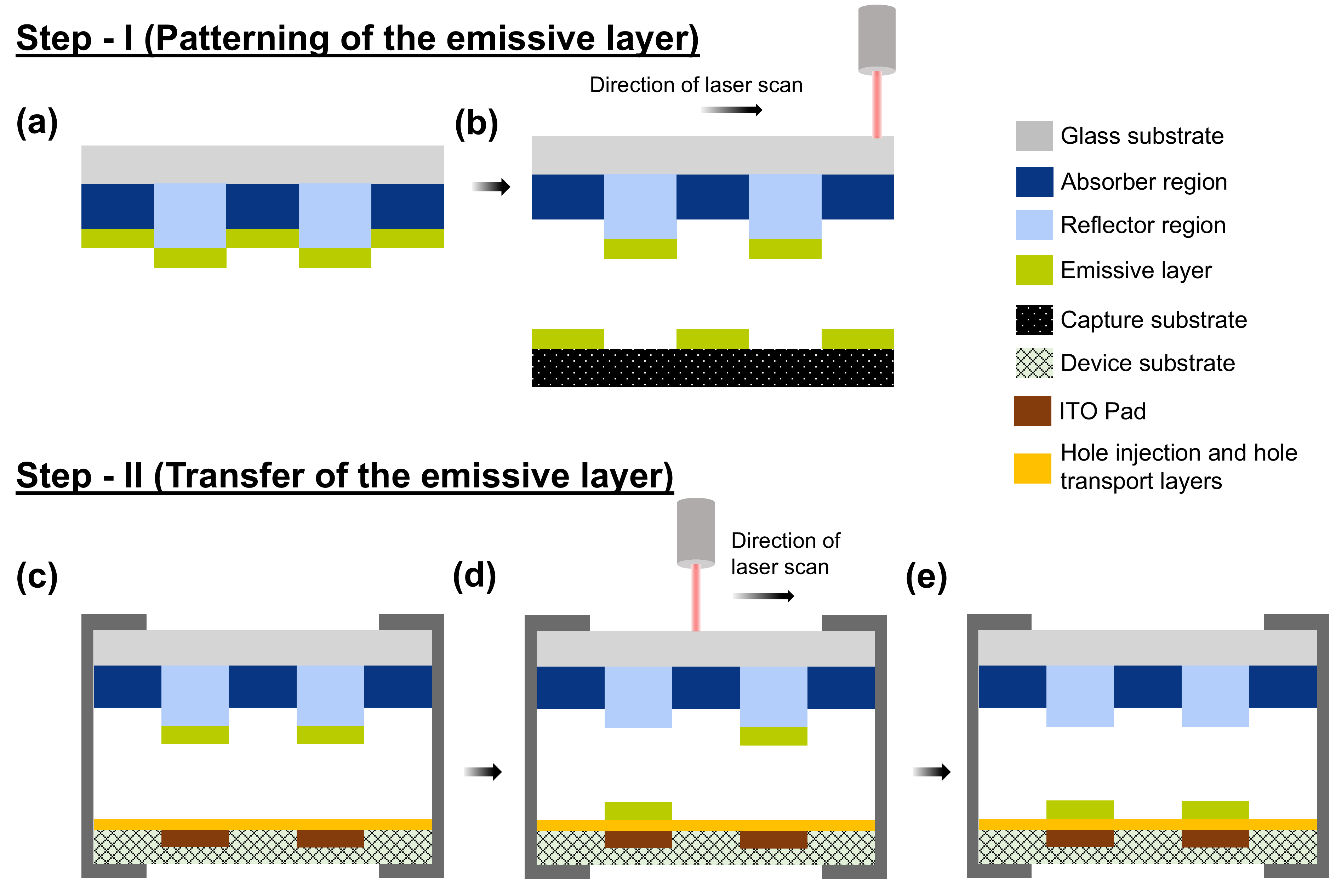}
	\caption{\textbf{The schematic of the patterning and transfer process} (a) The deposition of the emissive layer (light green) on the donor substrate, with absorber (dark blue) and reflector (light blue) regions. (b) Patterning of the emissive layer is achieved by laser irradiation, which sublimates the emitters from the absorber region to a capture substrate (which is a plain glass shown by a black background with white dots). (c) The precise alignment between the patterned donor substrate and the device substrate (which consists of the patterned indium tin oxide (ITO) pads and the hole injection and transport layers) is depicted. (d) The intermediate and (e) the final transfer of the patterned pixel to the device substrate are illustrated. The layer details in the schematic are shown in the legends in the top right of the panel.  }
	\label{fig:Schematic_PP_PT}
\end{figure}

Rapid transfer of organic layers, serving as pixels, was achieved via the laser-assisted close-space sublimation (LA-CSS) technique employing a specially patterned donor substrate, consisting of alternating absorber (high optical absorption) and reflector (low optical absorption) regions (Figure \ref{fig:Schematic_PP_PT}). After depositing the organic layer on the donor substrate, a laser with a linear intensity profile (perpendicular to the scan direction) is scanned across it, generating a temperature contrast between absorber and reflector regions. By tuning laser power and scan speed, the emitter above the absorber regions is selectively sublimated. A glass substrate is positioned in close proximity to the donor substrate, serving as a receiver (Figure \ref{fig:Schematic_PP_PT}(b)), capturing a replica of the sublimated pattern. The captured material can be reused, thereby reducing wastage. The patterned donor is then aligned with the device substrate containing common transport layers, and the same laser is used again, at higher power, to transfer the emitter pixels from the donor to the device substrate (Figure \ref{fig:Schematic_PP_PT}(c-e)). Accurate pixel transfer is governed mainly by the temperature difference between absorber and reflector regions and the donor–receiver spacing. Therefore, it is important to optimize the absorber and reflector material combination. A two step transfer process reduces the chance of colour mixing due to unintended material transfer on the device substrate.

\subsection{Choice of absorber and reflector combination for donor substrate}\label{subsec: choice of abs and ref}
The absorber and reflector pads consist of metals and metal oxides. An optimum combination was chosen from a pool of materials, M (M $\in \{\mathrm{Cr}, \mathrm{Ti}, \mathrm{W}, \mathrm{Mo}\}$) and  their oxides (M$_{x}$O$_{y}$)  based on the optical and thermal properties. The primary requirement of the absorber and reflector is to have a strong absorption at 940 nm, corresponding to the wavelength of the laser employed in the patterning and transfer experiment. This absorbed irradiation acts as the heat source. Furthermore, the material should exhibit high thermal conductivity to facilitate rapid heat transfer, thereby accelerating the evaporation of the organic layers from the donor substrate. In addition, a high specific heat capacity of the material is desirable. The objective was to identify an optimized combination of absorber and reflector materials such that a significant temperature difference is established between absorber and reflector under identical laser exposure conditions (interaction time and laser power).

\begin{figure}[ht!]
	\centering	
	\includegraphics[width=0.95\textwidth]{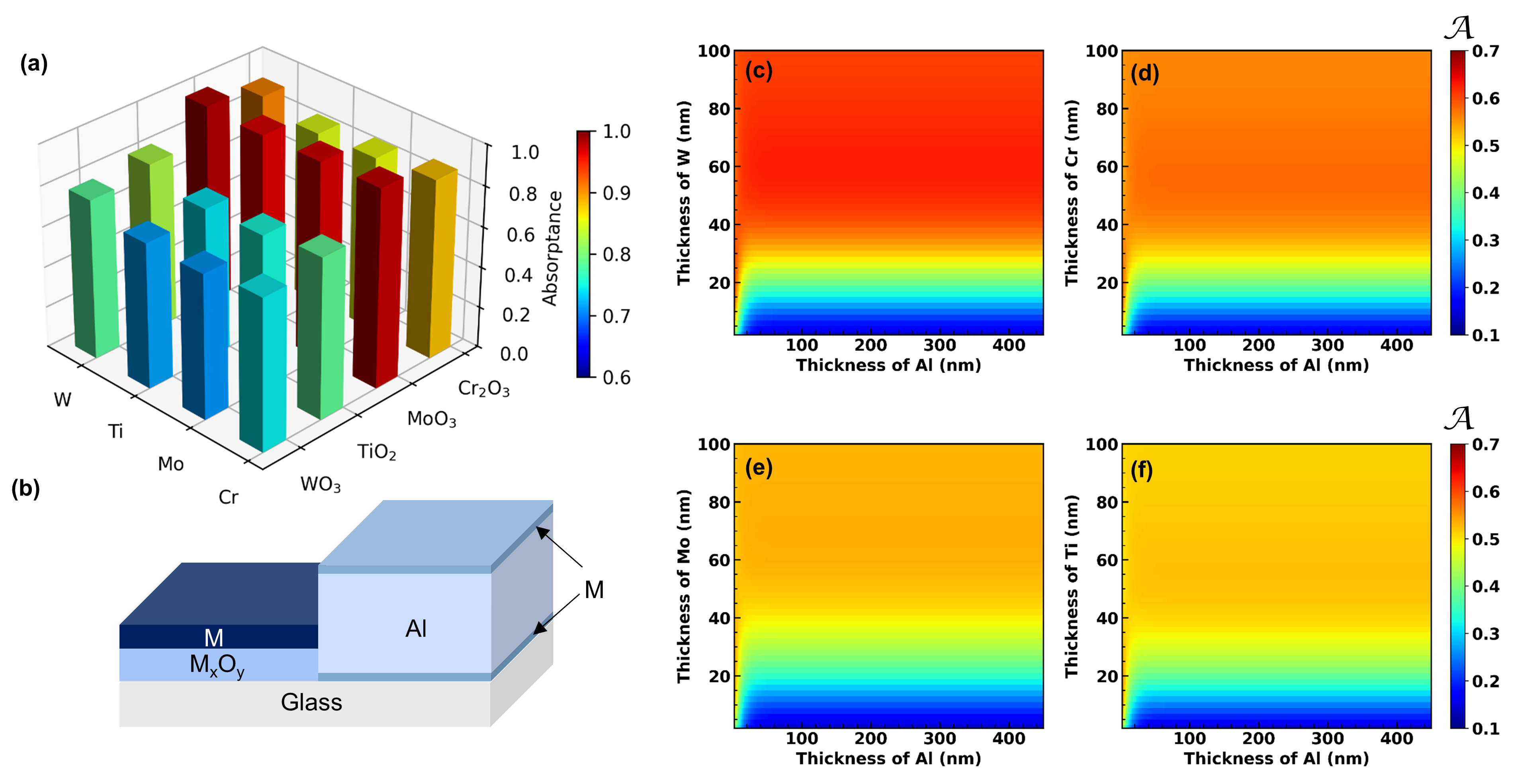}
	\caption{\textbf{Absorption of the absorber and reflector combinations and their schematic} (a) Comparison of maximum absorption for various combinations of metal and metal oxides. Each combination is represented by a rectangular bar, whose height denotes the corresponding absorptance. The absorptance magnitude can be inferred either from the bar height or from its colour, according to the colour scale shown in the colour bar on the right-hand side. (b) The schematic of the absorber and reflector structure. (c-f) The variation of calculated absorptance ($\mathcal{A}$) of the glass/M/Al/M (M $\in \{\mathrm{Cr}, \mathrm{Ti}, \mathrm{W}, \mathrm{Mo}\}$) combinations is shown for different thicknesses of Al and M  at 940 nm. The thickness of the top and bottom (c) W, (d) Cr, (e) Mo, and (f) Ti was same for a thickness of the Al.   }
	\label{fig:Abs_reflector_comparison_TMM}
\end{figure}

\noindent The absorption of the multilayer stacks (absorber and reflector) can be tuned by varying the thickness of the individual layers, which enables the temperature difference between the absorber and reflector. The absorption of the stacks was calculated using the transfer matrix method using layer thicknesses and their refractive indices, taken from the literature \cite{kulikova2020optical, sarkar2019hybridized, vos2016atomic, al2007optical, johnson1974optical, ordal1988optical, mathewson1971absolute}.

The structure of the absorber stack is glass/M$_x$O$_y$/M. Absorptance at 940 nm was calculated for various metals, oxides, and their thicknesses. The calculated absorptance ($\mathcal{A} = 1 - R - T$) ranged from 18.6\% to 98.0\% (Supplementary Figure \ref{fig:Absorptance_of_the_absorber_TMM}). For a given oxide thickness, absorption became independent of metal thickness once the thickness of the metal exceeded 30 $\pm$ 10 nm. The maximum absorptance oscillated with oxide thickness, with two bands observed near 90 $\pm$ 10 nm and 290-350 nm for all material combinations.

It is evident from Figure \ref{fig:Abs_reflector_comparison_TMM}(a) that for a specific oxide, M$_{x}$O$_{y}$/W has maximum absorption and M$_{x}$O$_{y}$/Ti has least absorption. Although the thermal conductivity of W is good (174 W/m.K), it has poor specific heat capacity (132 J/kg.K) \cite{haynes2016crc}. On the other hand, Ti has good heat capacity (532 J/kg.K), but thermal conductivity (20.9 W/m.K) is poor \cite{haynes2016crc}. Whereas, M$_{x}$O$_{y}$/Cr has second best absorption after M$_{x}$O$_{y}$/W. Cr has  a high specific heat capacity (448 J/kg·K) and a moderate thermal conductivity (87.8 W/m·K) \cite{haynes2016crc}.  Therefore, Cr was chosen as the metal in the absorber structure since it has good absorption along with better thermal properties to qualify as a good heat source.

From the Figure \ref{fig:Abs_reflector_comparison_TMM}(a), it is observed that for a specific metal, MoO$_{3}$ has maximum absorption, followed by Cr$_{2}$O$_{3}$. Although MoO$_{3}$ is clearly a good candidate from the perspective of absorption, the better thermal conductivity and specific heat of Cr$_{2}$O$_{3}$ (32.9 W/m.K and 822 J/Kg.K) \cite{gurevich2009thermodynamic, shackelford2000crc} than the MoO$_{3}$ (24.1 W/m.K and 521 J/Kg.K) \cite{kamoun2019nanostructured, haynes2016crc} make Cr$_{2}$O$_{3}$ a better candidate as a heat source. Therefore, Cr$_{2}$O$_{3}$/Cr was selected as the absorber material.

In the case of the reflector, the material combination with comparatively lower absorption is desirable. Aluminum (Al) and M (M $\in \{\mathrm{Cr}, \mathrm{Ti}, \mathrm{W}, \mathrm{Mo}\}$) combination was chosen due to the excellent thermal conductivity of the Al, which enables the better transfer of heat from one pixel to another. 
Al with higher specific heat would act as a heat source, while the metal would be used to tune the absorption accordingly. The absorption of various glass/M/Al/M (M $\in \{\mathrm{Cr}, \mathrm{Ti}, \mathrm{W}, \mathrm{Mo}\}$) combinations was calculated using TMM (shown in Figure \ref{fig:Abs_reflector_comparison_TMM}(c-f)). The absorptance became invariant for the Al thickness of more than 20 nm. It was maximum for metal thickness more than 40 nm. A maximum $\mathcal{A}$ of 0.62 was obtained in case of the glass/W/Al/W structure. Whereas the minimum $\mathcal{A}$ (0.52) was obtained for the glass/Ti/Al/Ti structure. For pattering of the substrate, it is desirable to have maximum contrast in absorptance between absorber and reflectors which will be reflected in their respective temperatures. Therefore, a glass/Ti/Al/Ti structure was chosen as the reflector.


\subsection{COMSOL simulation for the temperature profile of the patterned donor substrate}\label{subsec: COMSOL Simulation of temperature}

The temperature profile of the patterned donor substrate with an absorber and reflector was calculated using COMSOL multi-physics 5.6 \cite{bianco2008numerical}. The details of the method of the simulation  and the parameters used in the simulation are discussed in the supplementary  section \ref{section:COMSOL_Simulation_method}.  The absorber consists of glass/Cr$_{2}$O$_{3}$ (90 nm)/Cr (75 nm), and the reflector is composed of glass/Ti (50 nm)/Al (400 nm)/Ti (50 nm). The thickness of the layers was chosen based on TMM calculations.

The simulation was done in 2D (x-z plane of Figure \ref{fig:Abs_ref_structure}) as explained in supplementary section \ref{section:Abs_ref_structure}. The size of the absorber and reflector pad was 1500 $\mu$m.  The transient heat equation (equation \ref{equn:transient_heat_equation}) was solved to obtain the temperature as a function of the spatial coordinates (x, z) at various times, assuming thermally insulated boundaries and no radiation loss. The notation used in equation \ref{equn:transient_heat_equation} follows conventional definitions (details in supplementary section \ref{section:COMSOL_Simulation_method}).
\begin{equation}
\rho C_{p} \frac{\partial T}{\partial t} = k \nabla^{2} T + Q
\label{equn:transient_heat_equation}
\end{equation}

\begin{figure}[h!]
	\centering
	
	\includegraphics[width=0.98\textwidth]{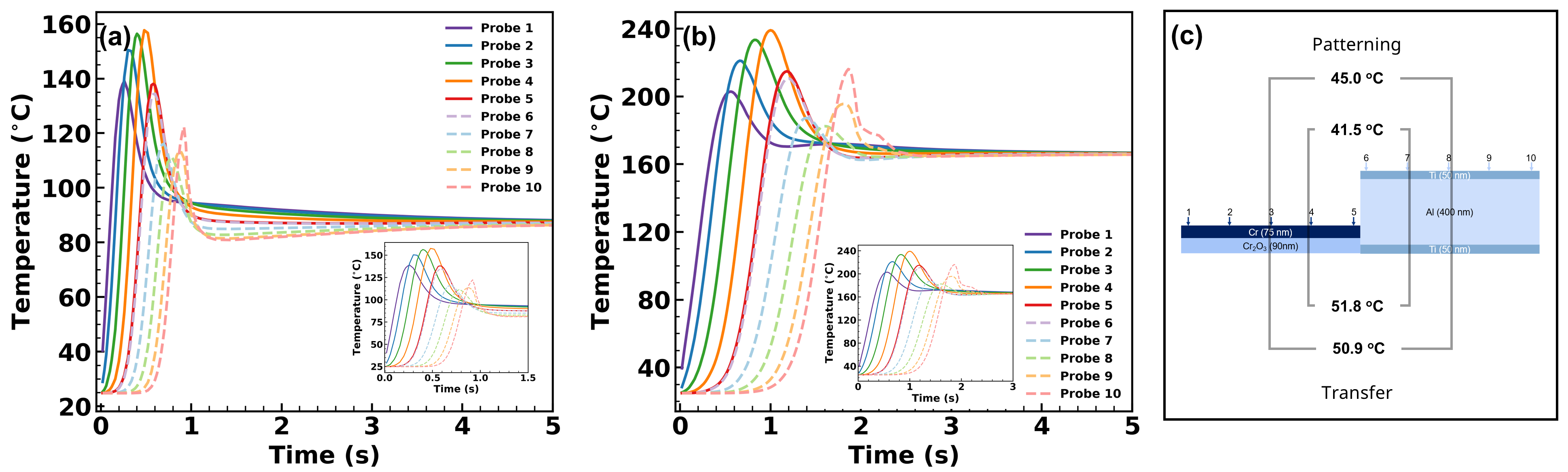}
	\caption{\textbf{Simulated transient temperature and temperature difference between absorber and reflector}. Transient temperature at various points on the absorber (solid bright line) and reflector (dashed broken line) under (a) patterning and (b) transfer conditions. The insets of figures (a) and (b) show the zoomed images. (c) The difference in temperature between the points on the absorber (left side) and reflector (right side) as obtained from the COMSOL simulation. The temperature difference between probe 3-8 and probe 4-7 for patterning and transfer conditions is shown at the top and bottom of the figure (c), respectively. }
	\label{fig:transient_temp_pastel_2_0_patterning_transfer_simulation}
\end{figure}

In simulation, five equidistant probes were placed on the surface of the absorber (probes 1-5) and the reflector (probes 6-10) to measure the transient temperature of the surface, as shown by the arrows in Figure \ref{fig:transient_temp_pastel_2_0_patterning_transfer_simulation}(c). Two different sets of laser parameters were used for patterning the substrate and for the transfer of the patterned organic layers.

For patterning, the average power density of the laser was 4.545 $\times$ 10$^{5}$ W/m$^{2}$ with a duty cycle of 75\% and scan speed of 4 mm/s. 
The maximum temperature on the absorber and reflector was 157 $^{\circ}$C and 111 $^{\circ}$C, respectively (Figure \ref{fig:transient_temp_pastel_2_0_patterning_transfer_simulation}(a)). The steady state temperature was 87 $^{\circ}$C. The difference between the maximum temperature of the absorber and reflector was 45 $^{\circ}$C between probe 3 and probe 8 and 41.5 $^{o}$C between probe 4 and probe 7 (Figure \ref{fig:transient_temp_pastel_2_0_patterning_transfer_simulation}(c)). 

For the transfer,  average power density of the laser was 4.773 $\times$ 10$^{5}$ W/m$^{2}$ with a duty cycle of 70\%. The scan speed was 2 mm/s. The evolution of the temperature at those points is shown in Figure \ref{fig:transient_temp_pastel_2_0_patterning_transfer_simulation}(b). The temperature at each point increased slowly to achieve the maximum values. Then all the points reached a steady state values. Since there was no loss, the steady state temperature showed a constant value with time. The maximum temperature (of 239 $^{\circ}$C) on the absorber was observed at the center (probe-3). At the center of the reflector, the temperature was 182.4 $^{\circ}$C (probe-8). The difference in the maximum temperature of absorber and reflector was around 50.9 $^{\circ}$C. Similar difference in temperature was observed in case of probe 4 and probe 7. The steady state temperature was 165 $^{\circ}$C.\\

\subsection{Compare the simulation results with the measured temperature}\label{subsec: Experimental temperature}

\begin{figure}[ht!]
	\centering	
	\includegraphics[width=0.9\textwidth]{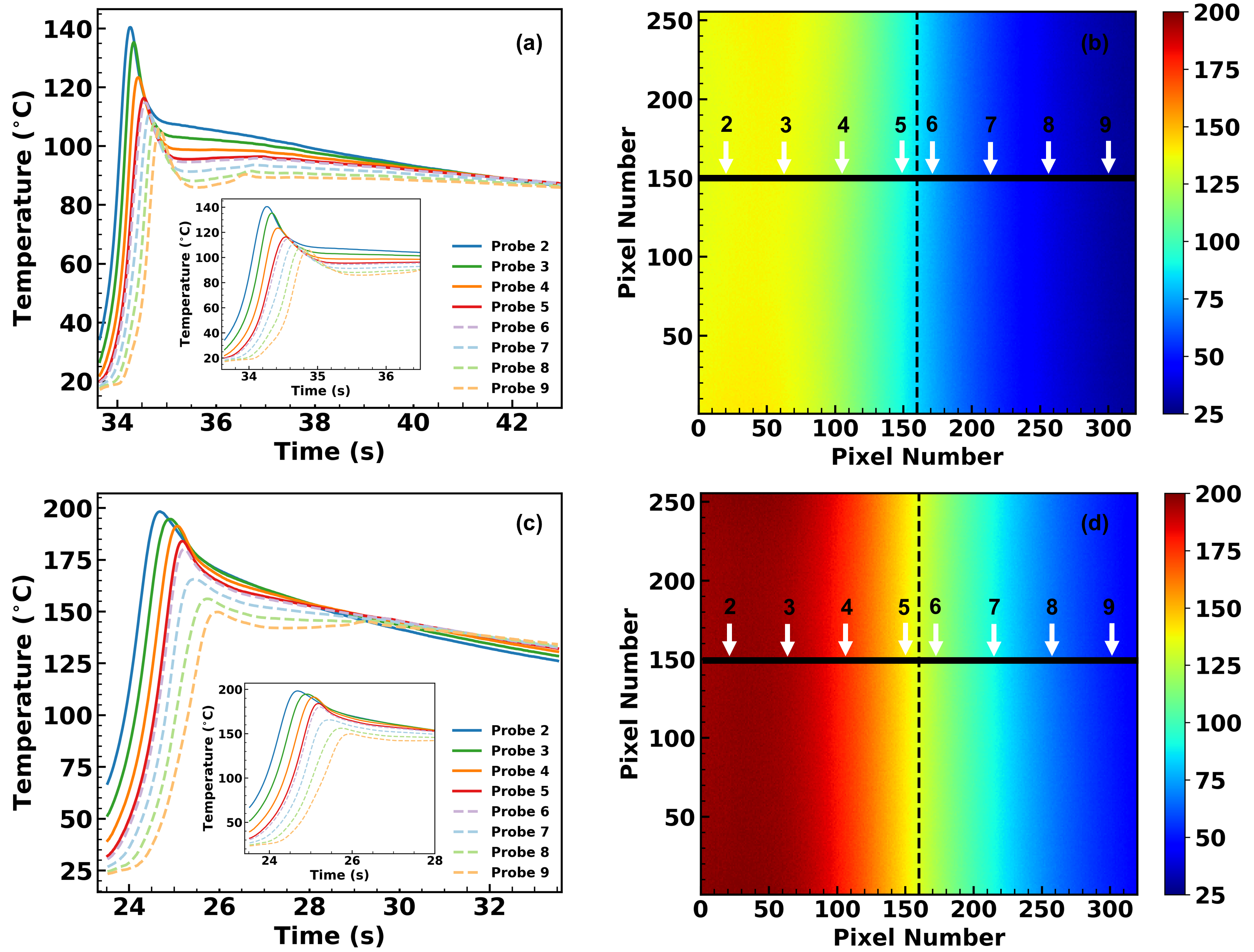}
	\caption{\textbf{Measured transient temperature for patterning and transfer condition}. Average transient temperature at various points on absorber (solid bright lines) and reflector (dashed light lines) under patterning (a) and transfer (c)  conditions as obtained from the experiment. The inset figures in (a) and (c) are the zoomed images, showing the variation of temperature for a small time after the laser interacted with the donor substrate.  Temperature profile of the surface as measured by the thermal camera for (b) patterning and (d) transfer condition. The junction of absorber and reflector is shown by the dotted vertical line. The solid black line showed the y position, where the average transient temperature was calculated. The white arrows represent the position of the probes, similar to the COMSOL simulation.   }
	\label{fig:transient_temp_pastel_2_0_patterning_transfer_experimental}
\end{figure}

The temperature profile of the surface of the absorber and reflector was measured using a thermal camera (details are in supplementary section \textcolor{blue}{ \ref{section:calibration_of_temperature}}), having a field of view of 2.67 mm $\times$ 2.14 mm. The thermal camera recorded the 2D temperature profile at a time interval. One of such frames is shown in Figure \ref{fig:transient_temp_pastel_2_0_patterning_transfer_experimental}(b). The junction of the absorber and reflector was shown by the vertical dotted line. The horizontal solid black line showed the top surface of the x-z plane where the average transient temperatures were probed. The position of the probe similar to the COMSOL simulation is shown by white arrows. The time stamp displayed in each frame is arbitrary and depends on the turn-on time of the thermal camera. Here, the time difference between the two frames is important. 

The temperatures were  measured using patterning parameters, described in section \ref{subsec: COMSOL Simulation of temperature}. The maximum temperature on the absorber was 140.3 $^{\circ}$C (Figure \ref{fig:transient_temp_pastel_2_0_patterning_transfer_experimental}(a)). At the center of the reflector, the maximum temperature  was 107.8 $^{\circ}$C. The steady-state temperature was 86.7 $\pm$ 0.7 $^{\circ}$C. The temperature decreased monotonically from absorber to reflector (from probe 2 - probe 9), in contrast to the simulation, where maximum temperature was observed at the center (probe 3) of the absorber. Then it decreased and settled down to a steady-state value, which was similar to the results obtained from the simulation.  This deviation could be due to the assumptions made in the simulation, such as no propagation of heat along the y direction and no loss of heat.  In a real system, the heat loss (by radiation and conduction) leads to the decrease in temperature, which was observed from the slope of the steady-state temperature in Figure \ref{fig:transient_temp_pastel_2_0_patterning_transfer_experimental}(a).  The maximum temperature at the center of the absorber was 135 $^{\circ}$C and 156.4 $^{\circ}$C as obtained from the experiment and simulation, respectively. At the center of the reflector, the maximum temperature obtained from experiment and simulation was 107.8  $^{\circ}$C and 112 $^{\circ}$C, respectively. The reduction in maximum temperature in experiment compared to the simulation in absorber and reflector are 13.7\% and 3.75\%, respectively. 

\noindent In Figure \ref{fig:transient_temp_pastel_2_0_patterning_transfer_experimental}(c), the temperature was measured for the transfer parameters as described in section \ref{subsec: COMSOL Simulation of temperature}.  In this case, a similar trend in temperature as the patterning temperature was observed. The temperature of the absorber was higher compared to the reflector side. The maximum temperature on the absorber was 198 $^{\circ}$C compared to the 239 $^{\circ}$C obtained in simulation. On the reflector pad, the maximum temperature was 156 $^{\circ}$C from experiment compared to 182.4 $^{\circ}$C obtained in the simulation. Since there was a gradient in transient temperature along the absorber and reflector both, the maximum temperature at the center of both as obtained from the simulation and experiment were compared. In both cases, the experimental temperatures  are 17.1\% and 13.8\% smaller compared to the temperature obtained from simulation.

\begin{figure}[ht!]
	\centering	
	\includegraphics[width=0.9\textwidth]{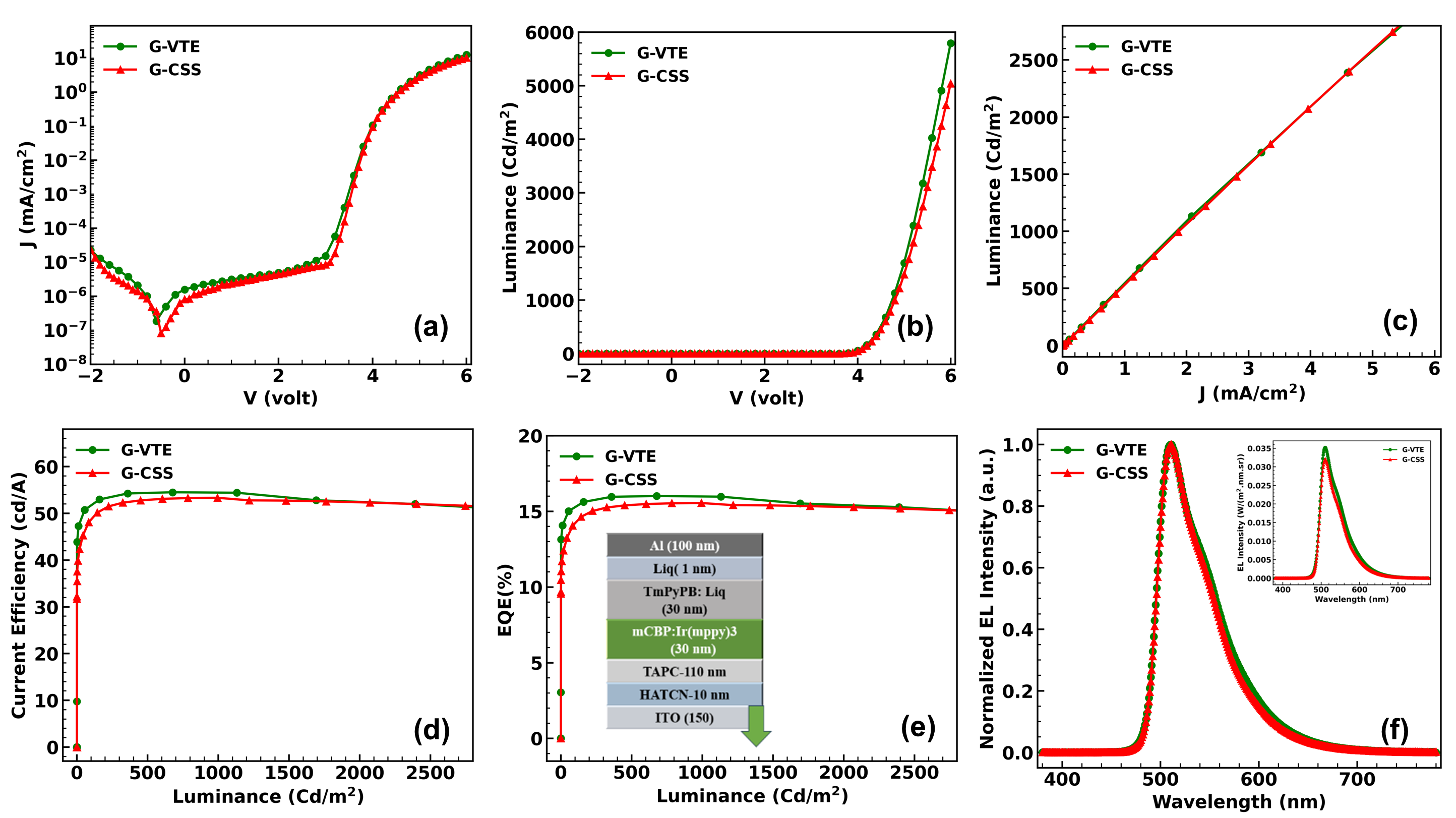}
	\caption{\textbf{The comparison of the device characteristics of the LA-CSS and VTE OLED}. The variation of (a) current density and (b) luminance as a function of applied voltage is shown. Variation of luminance with the current density is shown in figure (c). The change in (d) current efficiency and (e) EQE with luminance is plotted. The inset figure of (e) shows the device architecture used in this study. The bottom arrow shows the direction of collection of the emission. The normalized EL of the devices is shown in figure (f). The inset of (f) shows the absolute value of EL emission from the green VTE (solid circles with a green solid line) and green CSS (solid triangle with a red solid line) device.  }  
	\label{fig:CSS_vs_VTE_Device_results}
\end{figure}

\subsection{Laser-assisted CSS based OLED} \label{section: CSS_VTE_device} 
Using the two-step LA-CSS process, the emissive layer (EML) layer of the devices was deposited. The donor substrates were patterned with absorbers/reflectors where the active area was defined by the reflectors since the two-step process was used. The dimension of the active area was 3 mm $\times$ 3 mm. All other layers of the stack were deposited using the standard VTE process (details in section \ref{method:OLED_LACSS}). The EML consists of mCBP host and (Ir(mppy)$_{3}$) dopant of 5\% concentration. These devices are denoted as G-CSS devices. The response of these devices is compared with those grown entirely using VTE (G-VTE devices).

\noindent A comparison of the OLED characteristics of the G-CSS and the G-VTE devices is shown in the Figure \ref{fig:CSS_vs_VTE_Device_results}. The response of the two devices are nearly identical and are comparable to the best-reported devices \cite{guo2023low}. The device performance parameters are summarized in Table \ref{tab:VTE_CSS_OLED_Comparison}.  The current is not trap limited   at high current densities. It is probably due to the fact that the fabrication of the devices was performed without breaking vacuum in any step.  The similarity in luminance and current efficiency (Figure \ref{fig:CSS_vs_VTE_Device_results}(b), (c), (d)) for the devices grown by VTE and LA-CSS indicates that the mixing of the two molecules in the EML layer is homogeneous for both. This is supported by the electroluminescence spectrum, which shows no luminance from the host molecule (Figure \ref{fig:CSS_vs_VTE_Device_results}(f)). In case of segregation or inhomogeneous mixing, energy transfer from the host to guest can be suppressed leading to emission from the host. In a previous work, it was proposed that nearly equal molecular weight materials are required for homogenous mixing in the CSS process \cite{tam201912, dong2022low}. In the previous report, the CSS transfer was carried out using relatively slower Joule heating. Ir(mppy)$_{3}$ has a much higher molecular weight (696.86 g/mol) compared to mCBP  (484.59 g/mol). The fast laser transfer process leads to uniform mixing in this case. Hence, LA-CSS transfer has an inherent advantage over a slower Joule heating CSS transfer and no restrictions on molecular weights are required for uniform mixing of the molecules in the film. The similar response (luminance, current density, and EQE) of the devices (VTE and CSS) suggests a complete transfer of film from the donor substrate to the device for CSS, as any loss of film thickness would have led to a different device characteristic. This also suggests that there is no degradation of the material from the laser-assisted process. The laser is absorbed or reflected by the donor substrate, and hence, there is a very low probability of the laser interacting with the organic material coated on the donor substrate.

\begin{table}[h]
	\caption{VTE and LA-CSS based Green OLED device characteristics at 1000 cd m$^{-2}$ }\label{tab:VTE_CSS_OLED_Comparison}%
	\begin{tabular}{@{}llllllll@{}}
		\toprule
		Device & V$_{in} \footnotemark[1] (V)$  & V$_{on} \footnotemark[2] (V)$ & V$_{D} \footnotemark[3] (V)$ & $\eta_{CE}$ (cd.A$^{-1}$)  & EQE (\%) & $\lambda_{c}$ \footnotemark[4] (nm) & CIE  (x, y) \\
		\midrule
		G-VTE   & 3.4   & 3.6  & 4.75 & 54.4   & 15.9  & 510 & (0.285, 0.627)  \\
		G-CSS   & 3.5   & 3.65  & 4.8 & 53.3   & 15.5  & 511 & (0.278, 0.635)  \\
		
		\botrule
	\end{tabular}
\footnotetext[1]{V$_{in}$=Injection voltage}
\footnotetext[2]{V$_{on}$=Turn-on voltage}
\footnotetext[3]{V$_{D}$=Driving voltage}
\footnotetext[4]{$\lambda_{c}$=Peak position of emission wavelength }
\end{table}

\noindent The same study was conducted for a blue fluorescence OLED structure where the laser CSS-based OLEDs had the same response as those grown by VTE. The results are summarized in the supplementary section \ref{section:comparison_of_blue_OLED_LACSS_VTE}.  These results demonstrate the effectiveness of the LA-CSS method. 


\subsection{Pattering of micron-size features using the laser-assisted CSS process}

\begin{figure}[ht!]
	\centering	
	\includegraphics[width=0.9\textwidth]{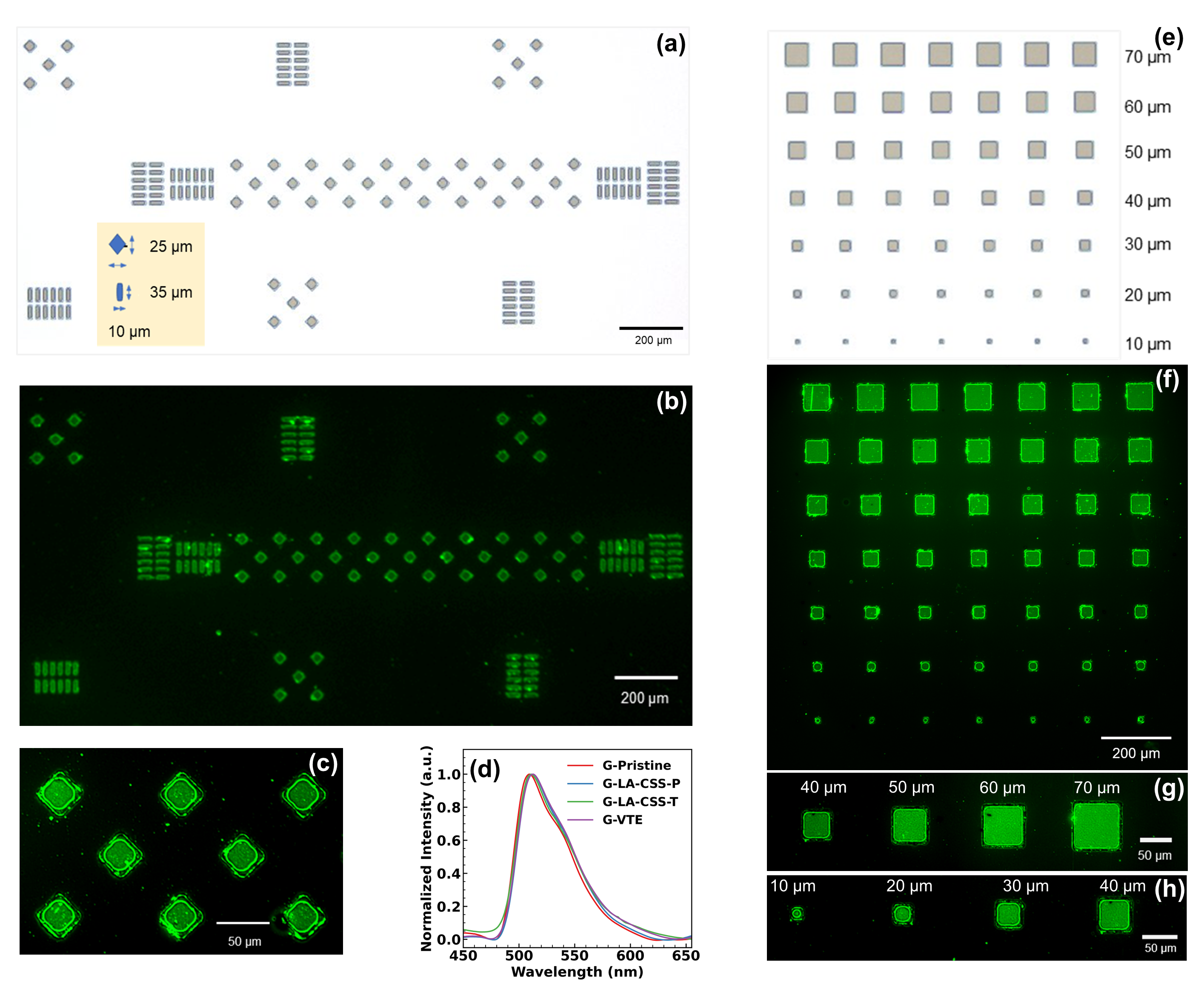}
	\caption{\textbf{Fluorescence microscopy images of the micro-pixels and their optical characteristics.} The test element group (TEG) (a) without an emissive layer and (b) with an emissive layer after patterning. The inset figure of (a) shows the dimensions of the reflectors. The zoomed pixels of TEG are displayed in figure (c). (d) The normalized PL spectra of the thin film of the pristine green emitter (Ir(mppy)$_{3}$) (represented by the solid red line) and its blend (mCBP:Ir(mppy)$_{3}$), which was fabricated by different methods. PL of the films, fabricated during LA-CSS patterning (LA-CSS-P) and LA-CSS transfer (LA-CSS-T), is shown by solid blue and green colors, whereas the PL of the film  by VTE is displayed by solid purple. The heat flow pattern (HFP)  (e) without an emissive layer and (f) with an emissive layer after patterning. The size of the squares on the  HFP is displayed on the right column. All the reflectors have the same size along a row. The zoomed pixels from the HFP are presented in figures (g) and (h). }
	\label{fig:Absorption_of_reflector_pastel_2_0_fluorescence_images}
\end{figure}

We generalize our method from macro-pixels to individual micro-pixels and examine the pattern fidelity of micron-scale features (10–70 $\mu$m) using LA-CSS. To assess transfer performance and limits in preserving microscale features, we use two substrates: a test element group (TEG) and heat flow patterns (HFP). The TEG includes 25 $\mu$m diamond structures and 35 $\mu$m × 10 $\mu$m rectangular patterns (Figure \ref{fig:Absorption_of_reflector_pastel_2_0_fluorescence_images}(a)), while the HFP consists of square reflectors from 10–70 $\mu$m in 10 $\mu$m steps (Figure \ref{fig:Absorption_of_reflector_pastel_2_0_fluorescence_images}(e)).

The green emitter Ir(mppy)$_3$, used as the OLED active layer, was used here as well.
Using optimized pattering parametres (average laser power density -
5.218 × 10$^{5}$ W/m$^{2}$ , 78\% duty cycle, 3 mm/s scan speed), organic material was removed only from the absorber region. The donor–receiver distance was $\sim$ 1.2 mm, which has little effect on feature size since the goal is solely removal of organics from the absorber region.

To evaluate the precision of the micron-sized transferred features, the donor substrate after transfer was characterized using an inverted fluorescence microscope (Olympus IX83). The fluorescence images show that the optimized parameters accurately patterned 25 $\mu$m and 35 $\mu$m TEG features (Figure \ref{fig:Absorption_of_reflector_pastel_2_0_fluorescence_images}(b) and (c)).

The same set of optimized parameters was employed to pattern the HFP. Features larger than 30 $\mu$m were transferred with high fidelity (Figure \ref{fig:Absorption_of_reflector_pastel_2_0_fluorescence_images}(f)–(h)), whereas 10 $\mu$m features exhibited edge distortions (Figure \ref{fig:Absorption_of_reflector_pastel_2_0_fluorescence_images}(h)). This distortion is likely attributable to lateral heat diffusion from the absorber into the adjacent reflector regions within the confined 10 $\mu$m feature areas, leading to unintended material removal from the reflector. 

Lateral heat conduction is determined by the donor substrate materials used in the dual-step laser-assisted CSS process as well as by the chosen transfer parameters. As discussed in section \ref{subsec: choice of abs and ref}, a sufficiently large contrast in thermal properties (e.g., absorptivity and thermal conductivity) between the absorber and reflector layers is required to ensure selective patterning. If this contrast is too small, lateral heat flow from the absorber into the reflector regions induces degradation and removal of organic material from the reflector during patterning, which manifests as feature distortion. A comparable deformation is also observed for the 20 $\mu$m features; however, the extent of deformation is reduced relative to the 10 $\mu$m features (Figure \ref{fig:Absorption_of_reflector_pastel_2_0_fluorescence_images}(h)).

After patterning the donor substrate, the residual materials on the reflector regions are completely transferred to the device substrate using the optimized transfer parameters (average laser power density: 5.118 $\times$ 10$^{5}$ W/m$^{2}$, 80\% duty cycle, and 1.5 mm/s scan speed). In this case, the distance between the donor substrate and the device substrate is 2 mm. This plays a prominent role in maintaining the feature dimension and fidelity.

The photoluminescence spectrum of LA-CSS transferred films, measured using a spectrofluorometer (JASCO FP-660), exhibits identical emission characteristics as patterned and VTE  films (Figure \ref{fig:Absorption_of_reflector_pastel_2_0_fluorescence_images}(d). This indicates uniform dopant distribution within the host matrix in both the patterned and pattern-transferred films, despite the substantial difference in molecular mass ($\sim$ 212 g/mol) between the host and dopant. In conventional Joule heating-based CSS transfer, maintaining a uniform host dopant mixture requires materials with similar molecular masses. As Joule heating is a slow process, transfer of dissimilar host-dopant material results in dopant segregation. In contrast, LA-CSS process is significantly faster, enabling rapid material transfer and thereby overcoming this molecular mass limitation.

\section{Conclusion}
We have demonstrated an LA-CSS method for the fabrication of OLEDs that exhibit device characteristics comparable to those produced by conventional VTE processes. This technique enables fast, maskless pixel patterning without damaging the organic emitters or compromising the homogeneity of the host–guest matrix, addressing key limitations of existing patterning approaches. By pre-patterning the donor substrate with absorber and reflector regions, we were able to generate a controlled temperature difference that is crucial for precise pixel formation. The temperature distribution obtained from COMSOL simulations showed good agreement with measurements from a thermal camera, validating our design strategy. Overall, LA-CSS opens up a promising, scalable route toward high-resolution OLED pixel fabrication and could facilitate more flexible and cost-effective manufacturing of next-generation advanced display technologies.


\section{Methods}
\subsection{OLED device fabrication using LA-CSS method} \label{method:OLED_LACSS}
The donor substrate and receiver glass substrates were cleaned using an optimized cleaning procedure, followed by spin-rinse drying (SRD) and UV–ozone surface treatment. Subsequently, the cleaned substrates were loaded into the load-lock chamber of a cluster system (Techno Blaze, Inc., Model 20190213). All device fabrication steps were performed within this cluster system, which comprises two chamber sections dedicated to vacuum thermal evaporation (VTE) and close-space sublimation (CSS), respectively.

The emissive layer, composed of mCBP (host) and 5\% Ir(mppy)$_{3}$ (dopant), was deposited onto the donor substrate via vacuum thermal evaporation (VTE), forming a film with a thickness of 30 nm. Following this, the donor substrate was patterned using LA-CSS  by selective removal of the emissive material. Inside the chamber, the donor substrate was mounted on the upper holder, whereas the glass substrate was mounted on the lower holder (Figure \ref{fig:Schematic_PP_PT}(b)). Laser patterning was carried out using a Hamamatsu Photonics laser system. Process parameters, including duty cycle, signal amplitude, scanning speed, and the gap between the substrates, govern the characteristics and quality of the resulting pattern.

After that using LA-CSS, the emissive layer was transferred to the indium tin oxide (ITO)-patterned device glass substrate, on which the hole injection layer (HIL) and hole transport layer (HTL) had been deposited via vacuum thermal evaporation (VTE). The donor substrate was mounted on the upper holder, while the device glass substrate was secured on the lower holder inside the CSS chamber. Prior to the transfer, the reflector region of the donor substrate was precisely aligned with the active area of the receiver (device) substrate. This alignment procedure ensures that material transfer is confined to the designated active region and prevents unintended material deposition outside the active region of the receiver substrate (Figure \ref{fig:Schematic_PP_PT}(c-e)).

Thus, the emission layer was integrated onto the device substrate using a two-step laser-assisted CSS process, which involves patterning and transfer. After the transfer of the emission layer, the remaining electron transport, injection layers and cathode were deposited using VTE.

\subsection{Characterization of the OLED devices} 
\label{method:OLED_LACSS_characterization}

The electrical and optical characteristics of the VTE and LA-CSS OLED devices were evaluated using a Current-Voltage-Luminance measurement setup (M6100 IVL system, McScience Inc., South Korea). This system incorporates a source meter to apply a voltage sweep across the specified range and a spectroradiometer to determine luminance, electroluminescence (EL) spectra, and color coordinates. Using the control software, the current density, luminance, EL spectra, CIE color coordinates, current efficiency, and EQE of the devices were recorded simultaneously. To ensure the reliability and reproducibility of the device performance, multiple devices with identical structures were fabricated and tested under the same measurement conditions. All measurements were conducted in a dark chamber to eliminate the influence of ambient light.



\bibliography{sn-bibliography}

\section{Acknowledgements}
The authors acknowledge the financial assistance provided by the Ministry of Electronics and Information Technology (MeitY), Govt. of India, Defence Research and Development Organization (DRDO), India, and Tata Sons Ltd., India. The authors thank Prof. Anjan Chakravorty for many useful discussions. We acknowledge HPCE, IIT Madras, for providing access to COMSOL 5.6.

\newpage
\newpage

\renewcommand{\thefigure}{S\arabic{figure}}
\renewcommand{\thetable}{S\arabic{table}}
\renewcommand{\thesection}{S\arabic{section}}
\setcounter{figure}{0}
\setcounter{table}{0}
\setcounter{equation}{0} 
\setcounter{enumi}{0} 
\setcounter{enumiv}{0} 
\setcounter{page}{1}
\setcounter{section}{1}
\setcounter{subsection}{0}

\begin{center}
	\textbf{\large{Supporting Information for}} \\
    \vspace{0.5 cm}
    \textbf{\large Mask-free fast patterning of organic light-emitting diode pixels using laser-assisted close-space sublimation} \\
    
    \vspace{0.5 cm}

    Subhamoy Sahoo$^{1}$, Jain Jose$^{2}$, Mani R$^{2}$, Arghya Saha$^{2}$, Kanimozhi V$^{2}$, Dhruvajyoti Barah$^{2}$, R. Bairava Ganesh$^{2}$, Amitava Majumdar$^{3}$, Jayeeta Bhattacharyya$^{1}$, G Rajeswaran$^{2}$, Debdutta Ray$^{2}$

    \vspace{0.2 cm}
    
     \textit{$^1$Department of Physics, Indian Institute of Technology Madras, Chennai, 600036, India}\\
     
     \textit{$^2$AMOLED Research Center, Department of Electrical Engineering, Indian Institute of Technology Madras, Chennai, 600036, India}\\
     
     \textit{$^3$Grantwood Technology Pvt. Ltd., New Delhi, 110019, India}\\

\end{center}

\subsection{Absorptance profile of the absorber for different combination of the metals and their oxides}
\begin{figure}[ht!]
	\centering	
	\includegraphics[width=0.9\textwidth]{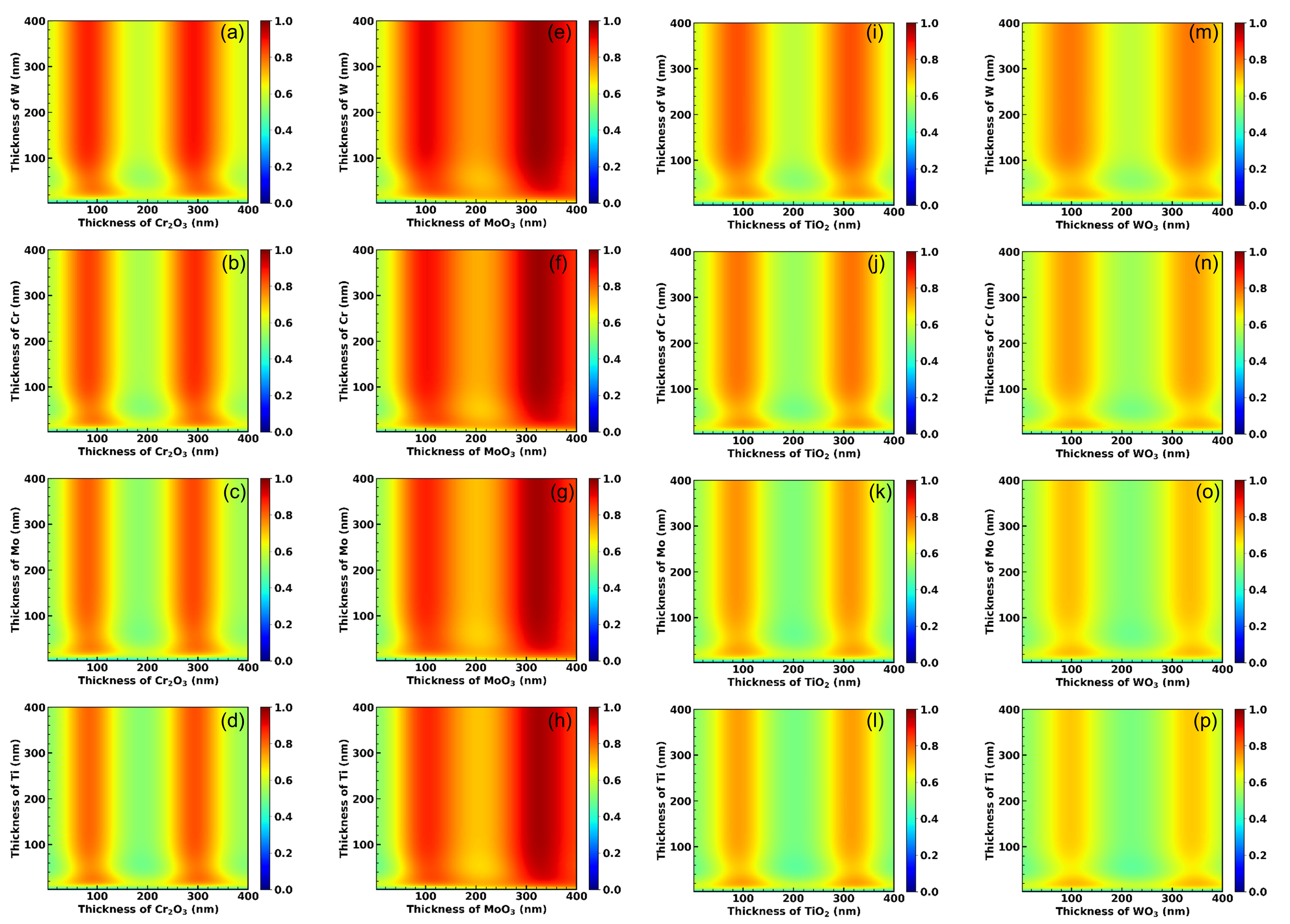}
	\caption{Absorptance of the absorber with different combinations of M (M $\in \{\mathrm{Cr}, \mathrm{Ti}, \mathrm{W}, \mathrm{Mo}\}$) and  their oxides (M$_{x}$O$_{y}$) }
	\label{fig:Absorptance_of_the_absorber_TMM}
\end{figure}

\noindent The calculated absorptance ($\mathcal{A} = 1 - R -T$ where R and T are the reflectance and transmittance) for  different combinations of M (M $\in \{\mathrm{Cr}, \mathrm{Ti}, \mathrm{W}, \mathrm{Mo}\}$) and  M$_{x}$O$_{y}$ as a function of their thicknesses is shown in the Figure \ref{fig:Absorptance_of_the_absorber_TMM}. The same colour scale is used for all the plots for the comparison.

\subsection{Methods and the parameters used in the COMSOL simulation} \label{section:COMSOL_Simulation_method}
The COMSOL simulation was performed in two steps. In the first step, the wave optics module was used to calculate the electric field profile along the absorber and reflector structure. This was then used to obtain the corresponding spatial distribution of absorbed power. In the second step, the transient temperature was calculated using the heat transfer module for a moving laser source. The shape of the laser is described in supplementary section \ref{section:shape_of_the_laser}. The simulation was done in 2D (x-z plane of  Figure \ref{fig:Abs_ref_structure}). The size of the absorber and reflector pad was equal (1500 $\mu$m).  The transient heat equation (equation \ref{equn:transient_heat_equation_suppl}) was solved to obtain the temperature as a function of the spatial coordinates (x, z) at various times. $\rho$, $C_{p}$, $k$, $T$, and $t$ represent density, specific heat at constant pressure, thermal conductivity of the material, temperature, and time, respectively. The term 'Q' on the right-hand side represents the heat source, that is, the absorbed laser power by the absorber and reflector. The initial temperature of the materials was taken to be room temperature, i.e., 298 K. The boundaries were assumed to be thermally insulated. Therefore, no heat was lost by conduction to the ambiance. Furthermore, it was assumed that there was no radiation loss. 
\begin{equation}
\rho C_{p} \frac{\partial T}{\partial t} = k \nabla^{2} T + Q
\label{equn:transient_heat_equation_suppl}
\end{equation}

\noindent The details of the refractive indices and parameters of the materials used for the simulation is given in table \ref{tab:refractive_index_and_thermal_parameters}.

\begin{table}[h!]
	\caption{Optical and thermal parameters used in the simulation  }
	\centering
	\begin{tabular}{|c|c|c|c|c|}
		\hline
		Details & Cr &	Cr$_{2}$O$_{3}$ &	Al  &	Ti \\ 
		\hline
		Refractive index (at 940 nm) &3.3574 &  1.29	 &	1.6737  &	3.3213 \\ 
		\hline
		Extinction coefficient (at 940 nm) & 3.5461 &	0.02 &	8.6085  &	3.9652 \\ 
		\hline
		c (J.Kg$^{-1}$.K$^{-1}$) & 448 &	822 &	921  &	532 \\ 
		\hline
		k (W.m$^{-1}$.K$^{-1}$) & 87.86 &	32.94 &	225.9  & 20.92 \\ 
		\hline
		$\rho$ (Kg.m$^{-3}$) & 7160 &	5250  &	2700  &	4500 \\ 
		\hline

	\end{tabular}
	\label{tab:refractive_index_and_thermal_parameters}
\end{table}


\subsection{Structure of the absorber-reflector used in the simulation} \label{section:Abs_ref_structure}
\begin{figure}[h!]
	\centering
	
	\includegraphics[width=0.9\textwidth]{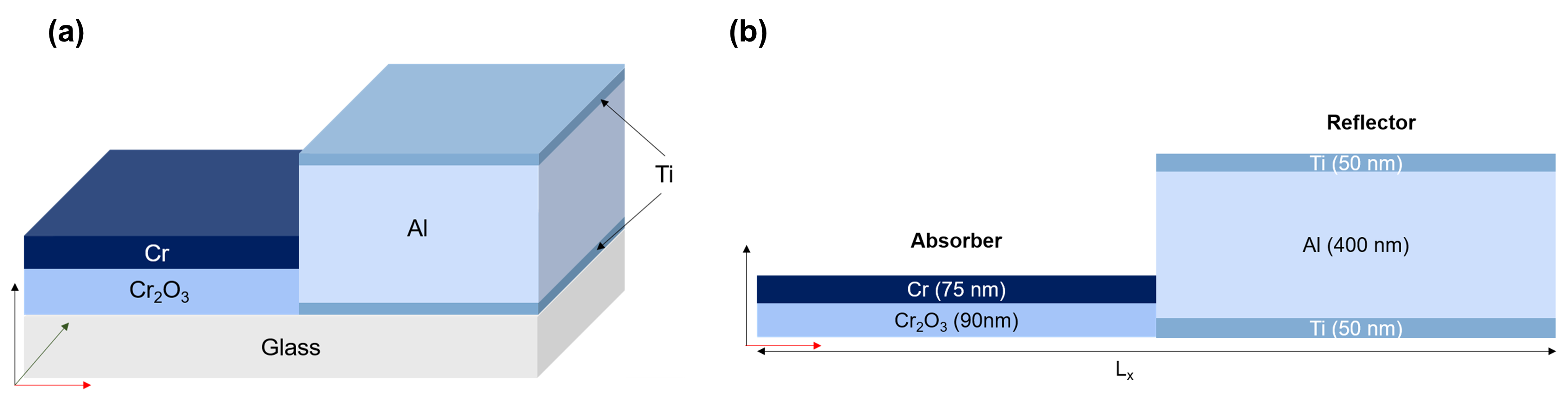}
	\caption{(a) The 3D structure of absorber and reflector. (b) the cross section of the structure used for the simulation. The x, y, and z axes are shown by red, green, and black arrows, respectively.  }
	\label{fig:Abs_ref_structure}
\end{figure}

The schematic of the combined absorber and reflector structure is shown in the Figure \ref{fig:Abs_ref_structure} (a). The x, y, and z axes are shown by red, green, and black arrows, respectively. The simulation was performed using the cross-sectional geometry of the original structure, as illustrated in the Figure \ref{fig:Abs_ref_structure} (b). The laser was moved from the start of the absorber towards the end of the reflector. Along the x direction, the intensity of the laser has a Gaussian distribution, and along y, the intensity is constant. Since the simulation was done for the cross section (x-z plane), the heat source in the simulation had a Gaussian distribution 
The moving source was defined by specifying the position of the Gaussian as a function of time by $x_{0} (t) = x_{0}(0) + \mathit{v}_{scan} t $. 

\subsection{Shape of the laser, used in the measurement} \label{section:shape_of_the_laser}
\begin{figure}[h!]
	\centering	
	\includegraphics[width=0.6\textwidth]{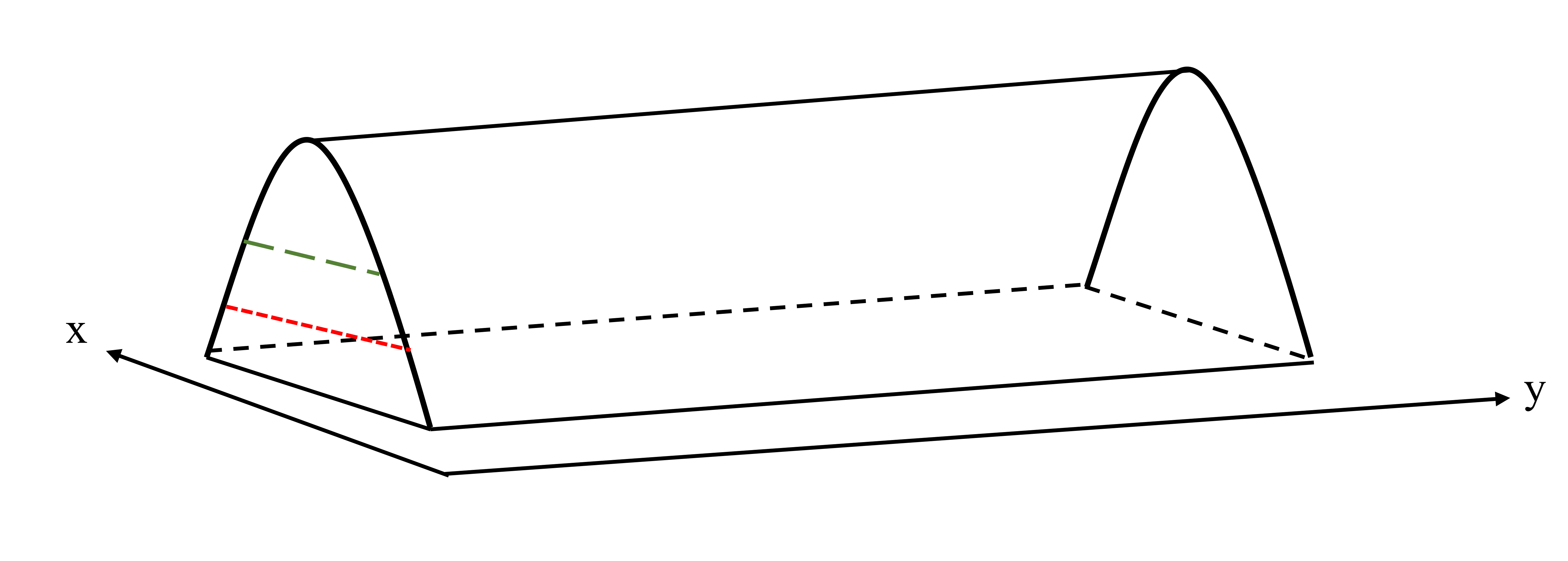}
	\caption{Intensity distribution of the laser used for the LA-CSS process. The intensity profile along the x and y directions has Gaussian and linear behavior, respectively. }
	\label{fig:Laser_shape}
\end{figure}

The laser (wavelength 940 $\pm$ 20 nm) used in this study has a Gaussian intensity profile along the \(x\)-direction, while it remains uniform along the \(y\)-direction. This is schematically depicted in Figure \ref{fig:Laser_shape}. The length along the y-axis was 80 mm, and the full width at half maxima (FWHM) along the x-direction was 1 mm. The laser can be operated in continuous wave (CW)  or pulsed mode by applying DC or AC voltage from  a function generator.

\subsection{Temperature measurement using thermal camera} \label{section:calibration_of_temperature}

\textbf{Experimental Setup and Instrumentation:}\\
Temperature measurements were performed using an infrared thermal imaging system (ViewOhre Imaging Co., Ltd., model XMCR32-SA0350-3XHT) positioned outside a vacuum chamber and aligned normal to the substrate surface. The chamber provided optical access for both laser irradiation and thermal imaging while minimizing convective heat loss.
Laser irradiation was introduced through an optical window using a fiber-coupled infrared laser system equipped with a focusing module (Hamamatsu, LE1287SP0LD lens unit). The laser beam was scanned across the substrate surface using a motorized linear translation stage (Chuo Seiki, QT-ADL1 controller with QT-AK interface), allowing precise control over the scanning velocity.
The infrared laser used for thermal excitation had a wavelength of 940 nm.
The laser beam was focused using the focusing optics. The same parameters were kept consistent during the thermal imaging measurements and were also used as input for the numerical simulations.\\

\noindent \textbf{Substrate Design and Surface Treatment:}\\
The substrate consisted of patterned absorber and reflector regions arranged in a 3 mm × 3 mm periodic geometry. Due to the difference in thermal coefficients between these regions, localized temperature variations were generated during laser irradiation.
As the substrate surface is inherently smooth and exhibits low emissivity, direct infrared measurements result in inaccurate temperature estimation. To mitigate this, a uniform coating of high-temperature black paint (Samurai KUROBUSHI, Hi-Temp Black) was applied to enhance surface emissivity.\\

\noindent \textbf{Emissivity Calibration:}\\
The emissivity of the coated substrate was calibrated using a thermocouple reference. The temperature measured by the thermal camera was correlated with thermocouple readings under controlled heating conditions. Based on this calibration, the effective emissivity of the coated surface was determined to be approximately 0.96, and this value was used in all thermal measurements.

\subsection{Comparison of device parameters of blue-OLED, fabricated using LA-CSS and VTE} \label{section:comparison_of_blue_OLED_LACSS_VTE}

\begin{figure}[h!]
	\centering	
	\includegraphics[width=0.9\textwidth]{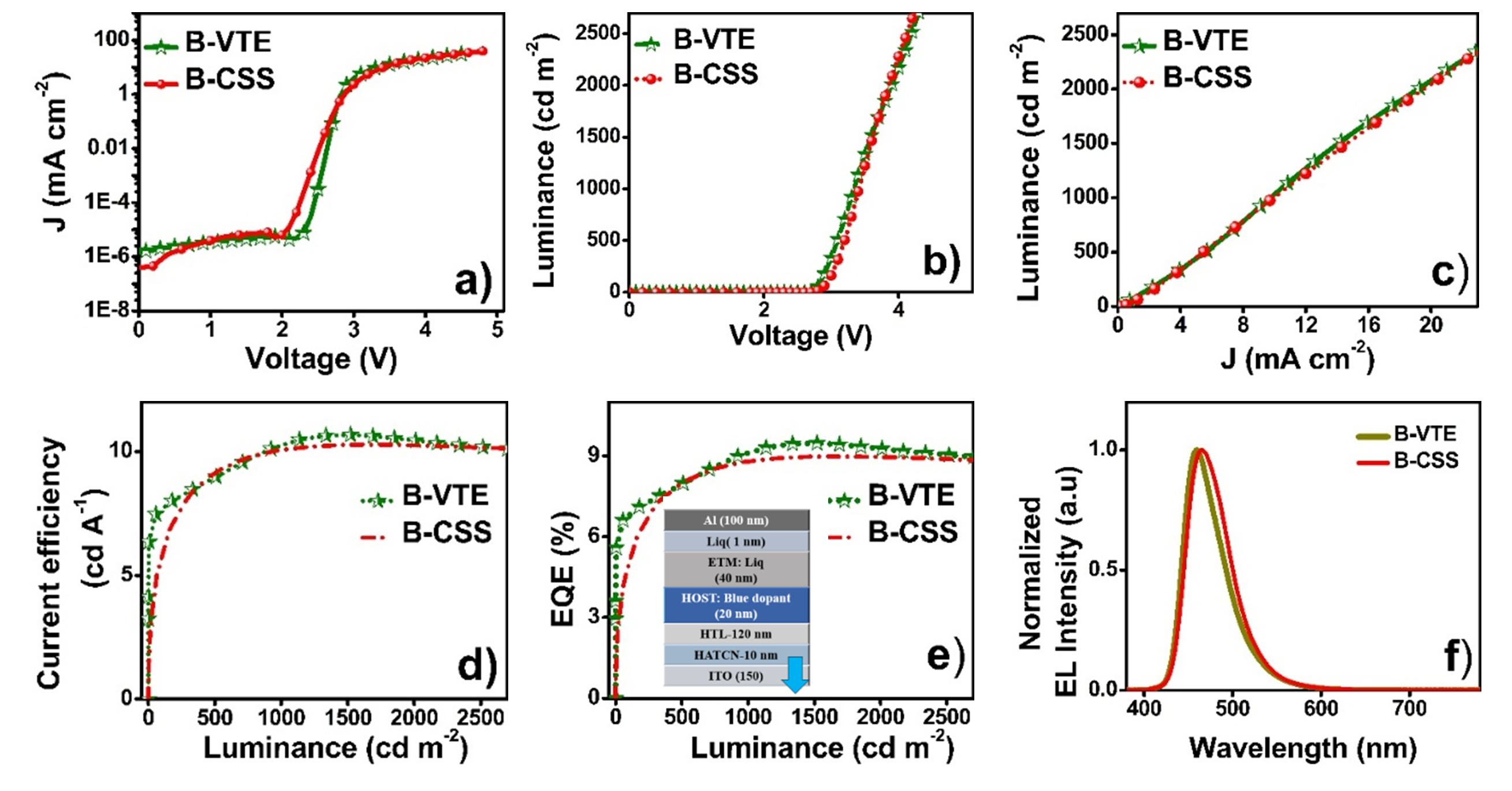}
	\caption{Characteristics of the blue OLED device, fabricated by the VTE and LA-CSS: The plot of (a) current density vs. voltage, (b) luminance vs. voltage, (c) luminance-current density relationship, (d) current efficiency vs. luminance characteristics, (e) external quantum efficiency vs. luminance plot, and (f) normalized electroluminance intensity. The schematic of the structure of the blue OLED device is shown in the inset of (e), where a proprietary blue host-dopant system, HTL, and ETL were used for the study.}
	\label{fig:comparison_of_blue_OLED_LACSS_VTE}
\end{figure}

\begin{table}[h!]
	\caption{VTE and LA-CSS based Blue OLED device characteristics at 1000 cd m$^{-2}$ }\label{tab:VTE_CSS_Blue_OLED_Comparison}%
	\begin{tabular}{@{}lllllll@{}}
		\toprule
		Device   & V$_{on} \footnotemark[1] (V)$ & V$_{D} \footnotemark[2] (V)$ & $\eta_{CE}$ \footnotemark[3] (cd.A$^{-1}$)  & EQE (\%) & $\lambda_{c}$ \footnotemark[4] (nm) & CIE  (x, y) \\
		\midrule
		B-VTE  & 2.65   & 3.4  & 10.4 & 9.3  & 460 & (0.14, 0.13)  \\
		B-CSS  & 2.65   & 3.4  & 10.0 & 8.7  & 466 & (0.13, 0.14)  \\
		
		\botrule
	\end{tabular}

\footnotetext[1]{V$_{on}$=Turn-on voltage}
\footnotetext[2]{V$_{D}$=Driving voltage}
\footnotetext[3]{$\eta_{CE}$=Current efficiency}
\footnotetext[4]{$\lambda_{c}$=Peak position of emission wavelength }
\end{table}

\newpage
.

\end{document}